# THE CONCEPT OF APPROPRIATION AS A HEURISTIC FOR CONCEPTUALISING THE RELATIONSHIP BETWEEN TECHNOLOGY, PEOPLE AND ORGANISATIONS

Paméla Baillette

CREGOR-Montpellier 2 University / GSCM-Montpellier Business School,

Montpellier, France

Tel/fax 04 68 66 22 63

pamelabaillette@yahoo.fr

Chris Kimble

University of York

York

UK

kimble@cs.york.ac.uk

*Abstract*

*The stated aim of this conference is to debate the continuing evolution of IS in businesses and other organisations. This paper seeks to contribute to this debate by exploring the concept of appropriation from a number of different epistemological, cultural and linguistic viewpoints to allow us to explore 'the black box' of appropriation and to gain a fuller understanding of the term. At the conceptual level, it will examine some of the different ways in which people have attempted to explain the relationship between the objective and concrete features of technology and the subjective and shifting nature of the people and organisation within which that technology is deployed. At the cultural and linguistic level the paper will examine the notion as it is found in the Francophone literature, where the term has a long and rich history, and the Anglophone literature where appropriation is seen as a rather more specialist term. The paper will conclude with some observations on the ongoing nature of the debate, the value of reading beyond the literature with which one is familiar and the rewards that come from exploring different historical (and linguistic) viewpoints.*

## 1. Introduction

This paper will explore the rich and multi-layered concept of appropriation as a way of examining the continuing evolution of ideas about how information systems relate to organisations and the people that work within them. It will do this by comparing and contrasting, differing historical, cultural and linguistic views of the concept of appropriation in general and the appropriation of technology in particular.

The paper consists of three main sections. Following this introductory section which provides some brief historical notes on the debate about the nature of relationship between technology, people and organisations, it outlines what might be called the traditional 'British' view of appropriation, rooted in the work of Marx (Marx 1977) and Braverman (Braverman 1974; Braverman 1998). This rather one-dimensional view of appropriation is then contrasted with the far richer view of concept found in the French literature. The final section of the paper concludes with some observations on the ongoing nature of the debate and the value of reading beyond the literature with which one is familiar.

**1.1  A brief history ...**

Probably the longest running debate in the field of Information Systems concerns the nature of the relationship between Information Systems and Organisations. In many ways, this debate itself is a continuation of the long-running agency-structure debate in Sociology and Psychology, although certain features of information systems do give this debate a new twist.

In the early days of information systems, the focus of the debate was largely on the effect that such systems had on managers (Kimble and McLoughlin 1995). For example, Leavitt and Whisler (1958) in an article entitled 'Management in the 1980s' made a number of predictions about the development of what they presciently called "information technology".

These predictions included that "information technology" would lead to:
(a) Top managers taking a far larger proportion of the innovating, planning and creative functions
(b) Fewer middle managers, with most of those who remained being routine technicians rather than thinkers
(c) Information systems allowing the top to control the middle, just as Taylorism allowed the middle to control the bottom, of the organisation

In contrast a later article (Applegate et al. 1988), written as a direct response to the article by Leavitt and Whisler 30 years earlier, argues that to merely react to new technology is a grossly inadequate response and suggests that managers should not simply respond to technological changes but should actively use them to shape the organisation. The argument is that it is the role of managers to decide how to develop and use IT and that they should not allow themselves to be driven by the technology.

These two positions, one based on notions of technological determinism where the technology itself plays the key role, and the other, which argues that people determine the effect of a technology not the other way round, dominated much of the thinking about information systems, people and organisations for some time. However, in the late 1980s and early 1990s new ideas, such as the work of Kling (1980; 1982) and ideas of Giddens and Orlikowski (Giddens 1984; Orlikowski and Robey 1991) began to open up a new, less deterministic view of this relationship.

While few doubted that there was a link between information technology / information systems and changes in organisational structures and the way in which people worked, there was little agreement, if any, about the underlying mechanism for those changes, or even what the changes were. For example, the paper referred to above (Kimble and McLoughlin 1995) comments on the way that the notion of technology 'impacting' on an organisation can itself be seen as biasing the debate in a particular direction. Consequently, the language that is used when discussing these issues is often as much a source of the problem as a solution to it.

It is into this background of disputed terms with contested meaning that we introduce another term - appropriation. The sections that follow will examine this term in more detail and from two distinct social and cultural perspectives: that of the English speaking (Anglophone) literature and that of the French speaking (Francophone) literature. It is our hope that this exploration can be used to throw some new light on to this old topic.

## 2   Braverman and Labour Process - a very 'British' view of appropriation

There is a historically specific notion of appropriation, and possibly one that is peculiar to Britain. It can be traced to the early works of Karl Marx and the later interpretation of those works by Braverman (1974; 1998).

## 2.1  Marx

Marx's notion of appropriation is most clearly seen in his work 'Economic and Philosophic Manuscripts of 1844'. The argument is essentially that, under the economic and political conditions of capitalism, the more the worker acquiesces to the demands of capital, the more they themselves become a commodity and the more they become alienated from their essential nature as a creative being.

> *"The product of labour is labour which has been embodied in an object, which has become material: it is the objectification of labour. Labour's realisation is its objectification ... this realisation of labour appears as a loss of realisation for the workers; objectification as loss of the object and bondage to it; appropriation as estrangement, as alienation ... so much does the appropriation of the object appear as estrangement, the more objects a worker produces, the less he can possess." (Marx 1977, p68)*

And later

> *"... thus the more the worker, by means of his labour, appropriates the external world, the more he deprives himself of the means of life." (Marx 1977, p 69)*

And again later

> *"... appropriation appears as estrangement, as alienation; and alienation appears as appropriation." (Marx 1977, p79)*

Thus for many the term appropriation became inseparable from the notion of 'taking something away' of 'the making of a thing into (private) property'. As Ashley and Plesch (2002) note:

> *"... 'appropriation' emphasizes the act of taking; it is understood to be 'active, subjective, and motivated' ... The fundamentally active nature of*

> *appropriation is manifest in its etymology, from the Latin verb appropriare, 'to make one's own,' a combination of 'ad', meaning 'to', with the notion of 'rendering to', and proprius, 'owned or personal'. Beyond the simple acknowledgment of borrowing or influence, what the concept of appropriation stresses is, above all, the motivation for the appropriation: to gain power." (Ashley and Plesch 2002, pp 2-3)*

## 2.2  Braverman

This theme of appropriation as loss and alienation was revived by Braverman's 1974 analysis of the labour process (Braverman 1974) where it was argued that work in the twentieth century, under the economic conditions of capitalism, represented an inevitable process of deskilling and subordination of labour.  This 'de-skilling Hypothesis' and Braverman's analysis of modern working practices was hugely influential at the time and, to some extent, continues to shape much of the research on 'Labour Process Theory' today (Tinkler 2002).

Like Marx, Braverman's approach was firmly rooted in the notion of the appropriation of the product of surplus labour, although Braverman's critique was directed specifically at Taylor's concept of Scientific Management.  For Braverman, Taylor represented the key to understanding the labour process, because he broke down the capitalist managerial imperative into its most basic elements.  Braverman argued that technology is not simply designed to improve production methods but to enhance managerial control of the production process.  The result is that workers are 'deskilled' as their tasks become routinised; their knowledge and skills are separated from them and compartmentalised and they become easier to isolate, manage and control (Noble and Lupton 1998).

Once again, the notion of appropriation is concerned primarily with the act of taking away, in this case taking away skills and knowledge from worker in order that they become easier to exploit.

## 2.3 Some alternative conceptions

The broadly Marxian conception of appropriation as loss influenced much of the work on the effects of 'new technology' in the 1980s and beyond. However, from the mid 1980s onward, a new view of appropriation began to emerge in the Anglophone literature; this became known as the Social Shaping of Technology (SST) movement.

In the UK, this movement emerged as a criticism of the perceived determinism of Braverman's followers. For example, Hughie MacKay argued:

> *"People are not merely malleable subjects who submit to the dictates of a technology: in their consumption, they are not passive dupes as suggested by crude theorists of ideology, but active, creative and expressive - albeit socially situated - subjects."* (MacKay 1992, p 698)

In an argument primarily directed at the determinist nature of much of the work up to that point he pointed out that much of this work had concentrated on the social forces underlying the creation of a technology and too little on the way in which technology can be actively appropriated by its users. For example, MacKay quoting Goodall (1983) argued:

> *"A new device merely opens a door: it does not compel one to enter. Technologies facilitate, they do not determine."* (MacKay 1992, p 701)

While acknowledging that the appropriation of a technology cannot be entirely separated from it design and development, the adherents of SST argue strongly for the role of the active and reflective subject in the way in which a technology 'becomes used'.

Meanwhile, in the US Rob Kling and Walt Scacchi (Kling and Scacchi 1980; Kling and Scacchi 1982) were also looking for new ways of thinking about the effect that technology might have on an organisation (and vice versa).

While Kling and Scacchi never used the term 'appropriation', it is clear that their interests focused on the links between what they termed "*computing and social life*". They state that most of the ideas in common use at that time (the early 1980s) were based on a highly simplified view that saw the effect a new technology had as simply a direct translation of certain specific technical attributes into equally specific social outcomes; Kling and Scacchi term these approaches 'discrete entity models'. They note that discrete entity models can be useful but warn that too often:

> *"Analysts who employ discrete entity models mistakenly assume that they are universal in their application." (Kling and Scacchi 1982, p5)*

As an alternative, Kling and Scacchi put forward their web model of computing to act as:

> *"... a conceptual vocabulary for describing and explaining the social events pertinent to computing development and use" (Kling and Scacchi 1982, p 10)*

Although Kling and Scacchi do not make this point explicitly, they portray discrete entity models as being primarily concerned with the technical development of systems: a situation where the subsequent use of a system is seen either as unproblematical or beyond the control of the analyst. Web models on the other hand focus on both the development and *use* of systems. This shift in emphasis not only allows the analyst to consider the effects of ongoing interactions but also how the effects of a system may change over time, such as when people 'learn' to use and exploit the characteristics of a new system.

While the ideas of Kling and Scacchi (1980; 1982), and later of McKay (1992), had some influence at the time, the broader concept of *appropriation as use* was never developed to the way it was in the French literature. Consequently, in the Anglophone literature at least, the notion of appropriation was left as a fairly specialist term in the vocabulary of structuration theorists such as Giddens (1984) and De Sanctis (1994).

## 3. The concept of appropriation in the French literature

From a Francophone point of view, the concept of appropriation does not pose any particular difficulties and is used frequently in everyday language (Bia-Figueiredo 2007). For example Carton et al. (2005), when discussing the appropriation of software tools, describe it simply as:

> *"... the processes by which individuals bring an unknown object into everyday use" (Carton et al. 2005)*

The idea of appropriation when used in conjunction with information and communication technologies (ICT) is generally seen as having positive connotations and is widely used in all types of organizations. In contrast to many of the 'Anglophone' views above, it is commonly held view that it is desirable to attempt to appropriate ICTs in order to make the best possible use of them.

However, if the positive connotation of appropriation of ICT does not seem to arouse controversy, a common description of it is much more difficult to find. Some define it as the outcome of a process, e.g. Proulx (2001b; 2001a; 2002) who considers the 'moment' of appropriation as the outcome of a sequential process, while others regard the term as applying to the process itself, for example, De Vaujany who states that:

> *"Appropriation is a long process that begins well before the use phase of the object and continues long after the onset of the first routinisation of use." (De Vaujany 2005, p 33)*

De Vaujany refers to the initial phase of this process as 'pre-appropriation'- the initial discussions and the evocation of the object, which is followed by a phase of 'original appropriation' where multiple socio-political or psycho-cognitive processes are activated within the organization, with the possibility of tensions, subsequently eased by the introduction of new routines; finally, the process of ends with the creation of a set of 'definitive' routines.

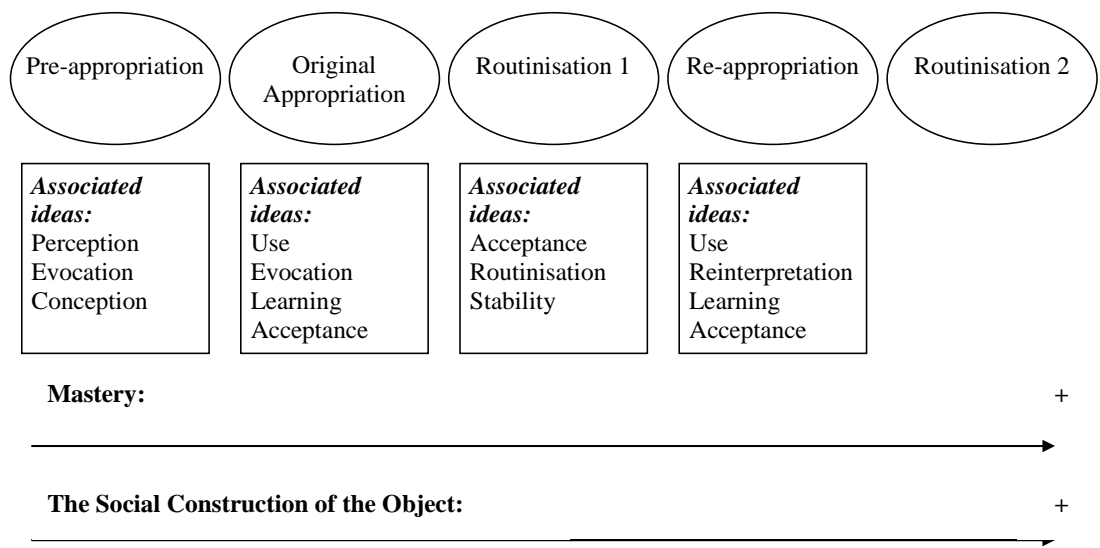

Figure 1 The process of appropriation by a collective (De Vaujany 2005, p 34)

Similarly, Massard (2007) also claims appropriation is a process: an organised phenomenon that evolves over time. This process is both individual, because it depends on the characteristics of each participant or actor, and collective, because it promotes the emergence of new structures within the organisation. Massard claims the term can cover three processes, each giving rise to a result called a 'state' of appropriation:

1)   A cognitive process
2)   A process of construction, in the sense of a technology
3)   A process of development of practice

Massard argues that the examination of these three processes will permit us to enter into the 'black box' of appropriation and gain a more complete understanding of the term. We will follow Massard's suggestion and briefly examine these three perspectives, illustrating each with some examples taken from the information systems field.

## 2.4 A cognitive process

A number of studies of appropriation share a common cognitive approach to the topic (e.g. Piaget and Inhelder 1992). In the Francophone literature on the science of language, the science of education and in ergonomics, appropriation is seen as a process that allows an individual to 'rebalance' (*rééquilibrer*) their internal cognitive structures following the disturbance in the environment, or to absorb new information from that environment. The term appropriation is often employed in these disciplines in preference to terms such as 'acquisition', adaptation', or 'learning' because it can encompass all of these concepts.

For example, Guillevic (1988) employed this 'cognitive' approach to appropriation in his study of technology transfer and the psychology of work. For him, appropriation was seen as a process of regulation following disruption from either external or internal factors such as factors peculiar to the individual or factors created by other individuals during the development or introduction of a new system of work.

## 2.5 A process of construction in the sense of a technology

In the literature on the sociology and psychology of work, appropriation is seen as the process by which an individual invests meaning and value in the use of a tool. The result of this process is characterised as the gap or difference in use between that imagined by the creators of the tool, and that which was put into effect by the end users, or by the different uses made of the same tool by different groups of users in the same context.

For example, when examining frameworks for the analysis of appropriation of technical objects, Perriault (1989) argued that the logic of the creator/conceiver of an object is to create a framework that prescribes the use of the object whereas the logic of the user, seen as an autonomous and creative actor, is to invent a use for that object. Thus, the process of the appropriation of a technical object or tool can be seen in the growing diversity of its use: the differences between the uses originally conceived of by the designer and those new uses invented by the users as the object is accepted into their everyday lives.

Similarly, Millerand (2003), in his study of electronic mail as a technology for learning among teachers and university researchers, observed that to appropriate technology is:

> "... to choose between a group of possible (actions) and to reinvent one's machine" (Millerand 2003. p 15)

The choice that Millerand refers to here is seen as a function of the meaning given to the use of the technology and the imagination of the user. The meanings given to the use are the representations and the values that the user invests in the use of a technique; the imagination of the user is the way the user organises his/her individual practice according to the different possibilities of use. The result of this process is an identity constructed from these meanings and practices.

### 2.6  A process of formation of practice

According to much of the Francophone management science literature, appropriation is the process by which routines of the organisation are constructed based on the properties of the technology. The result of this process is characterised by stability in terms of the structure of the organisation following the structural transformations.

For example, Prigent (1995) and Jouet (1993) both looked at the appropriation of systems of electronic messaging in an organisation and viewed appropriation as an intermediate process between representation and practice where:

> "Representation is the 'way of seeing' and practice is 'the act of doing'."
> (Prigent 1995)

This approach is very similar to that of De Sanctis and Poole (1994) who link Giddens' (1984) Structuration Theory and Ollman's (1971) Marxist analysis of appropriation to argue that information technology triggers adaptive structurational processes which, over time, lead to changes in the rules and resources that organisations use in social interaction.

## 2.7 Summary of Francophone views

The French literature has a very rich and diverse view of the concept of appropriation: it can to be taken to mean either a process or its outcome. Where it is considered a process, the term can be used to refer to both an individual process and a collective process. Proulx for example (2001b; 2001a; 2002) focuses on the behaviour of the 'human agent' or 'user' to characterize the appropriation of a technology as an individual process while other writers, such as Bianchi and Kouloumdjan (1985), Houzé (2000) and De Vaujany (2005), view it as a collective process. For example, Bianchi and Kouloumdjan claim that:

> *"... a group, a population, appropriate a given system of communications to in order to give them the keys (technical, economic, cultural, etc) to gain access to other users. In this way they implement the system in pursuit of their own goals." (Bianchi and Kouloumdjian 1985, p 145-146)*

If the authors do not always agree on a common definition of appropriation, most (e.g. Beaudry and Pinsonneault 1999; Isaac et al. 2006) recognize that it marks the integration of technology into the daily practice of users bringing benefits to both the individual and the organisation.

For example, Beaudry and Pinsonneault (1999) claim appropriation is a two dimensional process involving:

1. Integration of IT into one's work system (or task)
2. Integration of IT into one's work habits and routines

The effect this has on individual's performance depends on the degree of appropriation:

1. If the level of appropriation is high, then there are significant improvements in individual performance

2. If the level of appropriation is low then there is little or no positive effect on individual performance
3. If the level of appropriation nonexistent then there is a negative overall impact on individual performance

Even for those such as Proulx (2001a) who emphasise the moment of appropriation as the ultimate goal of some process, appropriation is seen as a positive, e.g. adaptation to the available technology in a way that is relevant to an individual's own career progression or progress. Thus Proulx (2001a), highlights three levels of appropriation that users can aspire to:

1. A minimum level of cognitive and technical mastery of the object or technical device
2. Integrating the social significance of the use of the technology in the daily life of the human agent
3. The possibility that creativity is facilitated by technology, that is that the use of technique allows the development of novelty in the life of the user

This section has highlighted the importance of the notion of appropriation of to Francophone researchers: if the introduction of a new technology is well managed, appropriation marks a successful integration of that technology into the daily practice of users.

## 3. Summary and conclusions

This paper has briefly reviewed the concept of appropriation as it appears in the Anglophone and Francophone literature; in particular, the way it is used in conjunction with the appropriation of (information) technology. Such a review must, of necessity, be partial and incomplete. It will also inevitably be open to suggestions that it takes a complex topic and turns it into two oversimplifies cultural stereotypes. The authors recognise these risks; they offer this paper not as a definitive study but more as a heuristic - a device that the reader might use to illuminate certain issues or topics that had not been noticed previously.

Having said our *mea culpas*, we will now attempt to summarise the two positions we have outlined and draw some conclusions about the value of the exercise.

Although the term clearly shares the same etymological roots, it seems reasonably clear that the same term represents two divergent concepts in the two bodies of literature. Broadly speaking, in the Anglophone literature appropriation is associated with something being taken from another: with one person (or group) gaining power at another expense. In contrast, appropriation in the Francophone literature is about taking something into one's self, without the overtones of depriving others by the act of doing so.

Applied to information systems, this results in the largely Anglophone view (e.g. Braverman 1974) that information systems can be seen as a way of appropriating the skills and knowledge of others for the benefit of a few. While this notion of appropriation is somewhat dated, it still shapes and influences the debate in the Anglophone literature - if only by omission. For example, in the Francophone literature authors such as Beaudry and Pinsonneault (1999) and Proulx (2002) have no problem discussing how users 'appropriate' information systems in the sense that they integrate such systems into their daily lives to the benefit to both the individual and the organisation. Is there a similar concept that could be deployed in the UK or US literature on the topic?

Finally, we conclude with a few general observations about the ongoing nature of the debate about technology and organisations and with some comments about the rewards that come from exploring different historical (and linguistic) viewpoints.

As we have seen, the debate has been part of IS literature for at least 50 years: probably as long as the notion of an information system has existed! Arguably, this can be seen as part of a longer debate about the nature of agency and structure in organisations. Whatever the truth of this, it is unlikely that the matter will be settled in the near future, if ever.

If this is the case, what value can we add by introducing a new term? The answer, to some extent, lies in the work of Edward Sapir (Sapir 1958) and Benjamin Lee Whorf

(Whorf 1956). According to what is now called the Sapir-Whorf hypothesis, our thinking is determined by language we used (i.e. linguistic determinism) and people who speak different languages perceive the world differently and are able to think and reason about it in different ways (known as linguistic relativity). Put at it simplest, this states:

> *"... the structure of a human being's language influences the manner in which he understands reality and behaves with respect to it"* (Carroll 1976, p 23)

The example most often given to illustrate this is the "Eskimos have 9 / 12 / 15 different words for snow" example, which states that because Eskimos have an extended vocabulary to describe snow, they can recognise feature of snow that others cannot and act accordingly.

The analogous argument for information systems is that by gaining a better understanding of richness of the concept of appropriation in the Francophone literature, we can gain new insights into the relationship between information systems and organisations that we would not have been able to express previously. This is much the same argument that Kling and Scacchi (1982) used when they attempted to introduce their 'Web Models' in the early 1980s which they described as:

> *"... a conceptual vocabulary for describing and explaining the social events pertinent to computing development and use"* (Kling and Scacchi 1982, p 10)

As we have seen, there are real differences in the way this concept is viewed in Anglophone and Francophone literature, so it is not so much as case of "*plus ça change, plus c'est la même chose*" but more a case of "*vive la différence*"!